\documentclass{ws-mpla}
\usepackage{feynmf}
\input slashed.sty
\usepackage{slashed}
\begin{document}


\catchline{}{}{}{}{}

\title{Multi Nucleon Mechanisms in Inclusive Subthreshold $K^+$ Production in $p$ $ ^{12}\mathrm{C}$ Collisions at 1 GeV\footnote{Supported in part by the Forschungszentrum J\"ulich (COSY)}}

\author{\footnotesize M. Dillig}

\address{Institute for Theoretical Physics III \footnote{preprint FAU-TP3-06/Nr. 02}, University of Erlangen-N\"urnberg,\\
Staudtstr. 7, D-91058 Erlangen, Germany\\
mdillig@theorie3.physik.uni-erlangen.de}

\author{C. Rothleitner}

\address{Max Planck Research Group, Institute of Optics, Information and Photonics, \newline University of Erlangen-N\"urnberg, \\
G\"unther-Scharowsky-Str. 1/Bau 24, D-91058 Erlangen, Germany\\
crothleitner@optik.uni-erlangen.de}

\maketitle


\begin{abstract}
We analyze recent inclusive data on $K^+$ production in proton induced $\Lambda$-multi nucleon knock out on $ ^{12}\mathrm{C}$. We describe collective effects among the incoming $p$ and bound nucleons in a $\pi, \rho$ rescattering model via the excitation of various baryon resonances. We find the dominance of $\Lambda N$ and $\Lambda NN$ final states for $K^+$ momenta below $200 MeV/c$, while for momenta above $400 MeV/c$ the data reflect a continuum with an unbound $\Lambda$ together with the bound residual nucleus $^{12}C$.

\end{abstract}

\ccode{PACS: 13.60 Le, 14.10 Cs, 24.40-n}

\section{Introduction}
One of the main goals of modern proton or electron accelerators at momentum transfers up to and around $1 GeV/c$ is to investigate the short range structure of the nucleon-nucleon interaction in the two-nucleon systems\cite{Machner}. Beyond that, reactions involving a large momentum transfer on nuclei with $A > 2$, allow to test multi-nucleon mechanisms or collective effects in the nuclear medium. This is most clearly seen in inclusive reactions, where nucleons are knocked out into the continuum, such as $pA \rightarrow NN\dotsb NA^\star$ (with $A^\star$ being the residual nucleus). Investigations to shed light on multi nucleon mechanisms have been started at COSY. In a recent paper\cite{Koptev} data are presented for the invariant cross section of the inclusive reaction $p(1 GeV/c)\; ^{12}\mathrm{C\rightarrow K^+X}$ as a function of the $K^+$ momentum from $150 MeV/c$ to above $500 MeV/c$ (close to the exclusive production of $^{13}_{\Lambda}\mathrm{C}$). For the kinematics far below the free $NN$ threshold the data above $400 MeV/c$ should reflect according to the estimates of the authors a high degree of collectivity among 6-7 bound nucleons, which are emitted as a cluster into the continuum .\\
\newline \indent However, the conclusions above have to be taken with care; the estimate is based on the assumption that the $\Lambda K^+$ system is produced in the collision of the projectile with a single heavy nucleon cluster, which is subsequently emitted into the continuum. We approach the data differently; experience from other near-exclusive high momentum transfer processes suggest the following microscopic picture: the dynamical collectivity (beyond the pure center-of-mass correlation of all nucleons) is an interplay of strength of the $NN$ rescattering amplitude, phase space and momentum sharing constraints among the bound nucleons. At low $K^+$ momenta the characteristic momentum transfer onto the target of $q\sim\sqrt{3}M\sim1500MeV/c$ ($M$ is the mass of the nucleon) is shared among typically 1-2 nucleons in the continuum: such a distribution combines maximal phase space with momentum sharing on each bound state wave function near the peak of the momentum distribution of $p$-shell momentum density of around $200MeV/c$. A higher degree of collectivity is suppressed: little gain in momentum sharing is overbalanced by decreasing rescattering contributions of higher order reflecting the strength of the elementary (off shell) $NN$ amplitude.

\section{The model}
We substantiate these heuristic arguments by a microscopic calculation: we evaluate the full transition amplitude with $N$ collectively cooperating nucleons in
$ ^{12}\mathrm{C}$ and $M$ nucleons in the continuum

\begin{align}
T(N,M)=\sum_{i>j}\langle K_{\Lambda},K_1\dotsc K_M, K_A&(M+1,\dotsc ,N,\dotsc ,12)\nonumber\\
& \times \vert V(i,j,K_k)\vert K,A(1,\dotsc ,12)\rangle
\label{glg1}
\end{align}

\noindent where we model $ ^{12}\mathrm{C}$ as s and p shell nucleons in an harmonic oscillator potential\cite{Vries} (which yields multiple overlap integrals, which can be performed analytically). We model the single and multiple rescattering mechanisms via $\pi$ and $\rho$ exchange; furthermore we assume that the $K^+\Lambda$ system is produced directly from the incoming proton via the baryon resonances $N^{\star}$(1650), $N^{\star}$(1710) and $N^{\star}$(1720)\cite{Groom,Sibirtsev}, whereas rescattering on the nucleons in $ ^{12}\mathrm{C}$ is dominated by the excitation of the $\Delta$(1232) isobar.\\

In this picture the static vertex functions for the $\pi$-vertices are given in an obvious notation by
\begin{align}
L_{\pi NN^{\star}} & = g_{\pi NN^{\star}}\underline{\tau}\underline{\phi}_{pi}\nonumber\\
L_{\pi N\Delta} & = \frac{f_{\pi N\Delta}}{m_{\pi}}\underline{S}^+\underline{q_{\pi}}\underline{T}^+\underline{\phi_{\pi}}
\label{glg2}
\end{align}

\noindent for the coupling to the s-wave $N^{\star}$(1650) and the $\Delta$-isobar, respectively,
(for the additional p-wave $N^{\star}$ resonances the $\pi$-coupling is as to the $\Delta$(1232)
isobar with appropriate coupling constants and $\underline{S}^{+} \rightarrow \underline{\sigma},\;
\underline{T}^{+} \rightarrow \underline{\tau} $), while the exchange of the $\rho$-meson reads

\begin{align}
L_{\rho NN^{\star}} &=g_{\rho NN^{\star}}\underline{\varepsilon}\underline{\sigma} \,  \underline{\tau}
\underline{\phi}_{\rho}\nonumber\\
L_{\rho N\Delta} &= \frac{f_{\rho N\Delta}}{m_{\rho}}\underline{\varepsilon}(\underline{S}^+\times\underline{q}_{\rho})\underline{T}^+\underline{\phi}_{\rho}
\label{glg3}
\end{align}

\noindent and correspondingly to the other $N^{\star}$-resonances included. The various coupling constants are taken for the $\pi NN$ and $\rho NN$ vertices from the Bonn potential\cite{Machleidt}; for the $K^+\Lambda N^{\star},\; \pi NN^{\star}$ and $\pi N\Delta$ vertices, the coupling constants are extracted from the corresponding partial decay widths into the $\pi$ and $K$ channel\cite{Groom} or from the constituent quark model for the $\rho$-meson\cite{Riska}. Then, as a typical example, double pion rescattering reads for the interaction in momentum space for the excitation of the \newline $S_{11} N^{\star}$(1650) resonance at the $K^+N\Lambda$ vertex
\begin{align}
V(N_1,N_2,N_3,\Lambda ,K^+) &=\frac{g_{K^+\Lambda N^{\star} }g_{\pi NN^{\star}}}{\underline{q}_1^2 +m_{\pi}^2-\omega_1^2}\frac{\left( \frac{f_{\pi N\Delta}}{m_{\pi}}\right) ^2 \underline{S}_2\underline{q}_2\underline{S}^+_2\underline{q}_1} {\underline{q}_2^2+m_{\pi}^2-\omega_2^2}\nonumber\\
& \quad \times \frac{\left( \frac{f_{\pi N\Delta}}{m_{\pi}}\right)^2 \underline{S}_3\underline{q}_3\underline{S}_3^+\underline{q}_2} {\underline{q}_3^2+m_{\pi}^2-\omega_3^2}(\underline{\tau}_1\underline{T}_2^+)(\underline{T}_2\underline{T}_3^+)
\label{glg4}
\end{align}

\noindent where the energy transfers$\omega _i$ are fixed from energy conservation at each vertex. Performing the Fourier transform to coordinate space upon integration over $\underline{q}_i$ we find characteristically

\begin{equation}
\int \frac{e^{\text{i}\underline{q}\underline{r}}}{\underline{q}^2+m_{\pi}^2-\omega^2}
d\underline{q}\sim\left\{
\begin{split}
 e^{-\sqrt{m_{\pi}^2-\omega ^2}r}/r & \quad\text{for} \quad  m_{\pi}\geq\omega \\
(\cos( \sqrt{\omega ^2-m_{\pi}^2} \,r)+\sin ( \sqrt{\omega ^2-m_{\pi}}\,r))/r  & \quad \text{for} \quad  \omega > m_{\pi}
\end{split}
\right.
\end{equation}

\noindent For an analytical result the various meson propagators are expanded in a sum of Gaussians; then multiple integrals in the total transition amplitude in coordinate space are easily evaluated with the standard formula\cite{Gradshteyn}

\begin{align}
\int Y_{lm}(\hat{r})r^n e^{-ar^2+\text{i}b\underline{q}\underline{r}}d\underline{r} & =\pi^{\frac{3}{2}}( \frac{\text{i} b}{2}) ^l\frac{\Gamma( \frac{n+l+3}{2})} {\Gamma(l+\frac{3}{2})}\nonumber\\
& \quad\times(q^2)^{\frac{l}{2}}Y_{lm}(\underline{\hat{q}})e^{-\frac{b^2\underline{q}^2}{4a}}
_1F\left( \frac{l-n}{2},\; l+\frac{3}{2},\; \frac{b^2\underline{q}^2}{4a}\right)
\label{glg6}
\end{align}

\noindent where $\Gamma$ and $F$ denote the Gamma and the confluent hypergeometric function of 1st kind, respectively \cite{Abramowitz}. Finally, in calculating the cross section, the $\Lambda$-multi nucleon phase space is included in a non relativistic approximation\cite{Shyam}.

\section{Results and discussion}
Within the model presented in the previous section we calculate the momentum distribution of the $K^+$ meson in $\mathrm{p+\, ^{12} C\rightarrow K^+ +X}$ at a proton momentum of $1GeV/c$ (in the lab. system). In the calculation we include $\pi$ and $\rho$ rescattering to the 6th order, allowing the knock out of up to six nucleons into the continuum. $^{12}\mathrm{C}$ is represented as the canonical superposition of s and p states in a single particle model with harmonic oscillator wave functions; the size parameter for these states is taken from electron scattering as $a=1.67fm$\cite{Vries}.\\

A typical result of our calculation is presented with a comparison to the data from ref. \cite{Koptev} in Fig. 1.
\begin{figure}[th]
\begin{center}
\includegraphics[scale=0.4,angle=-90]{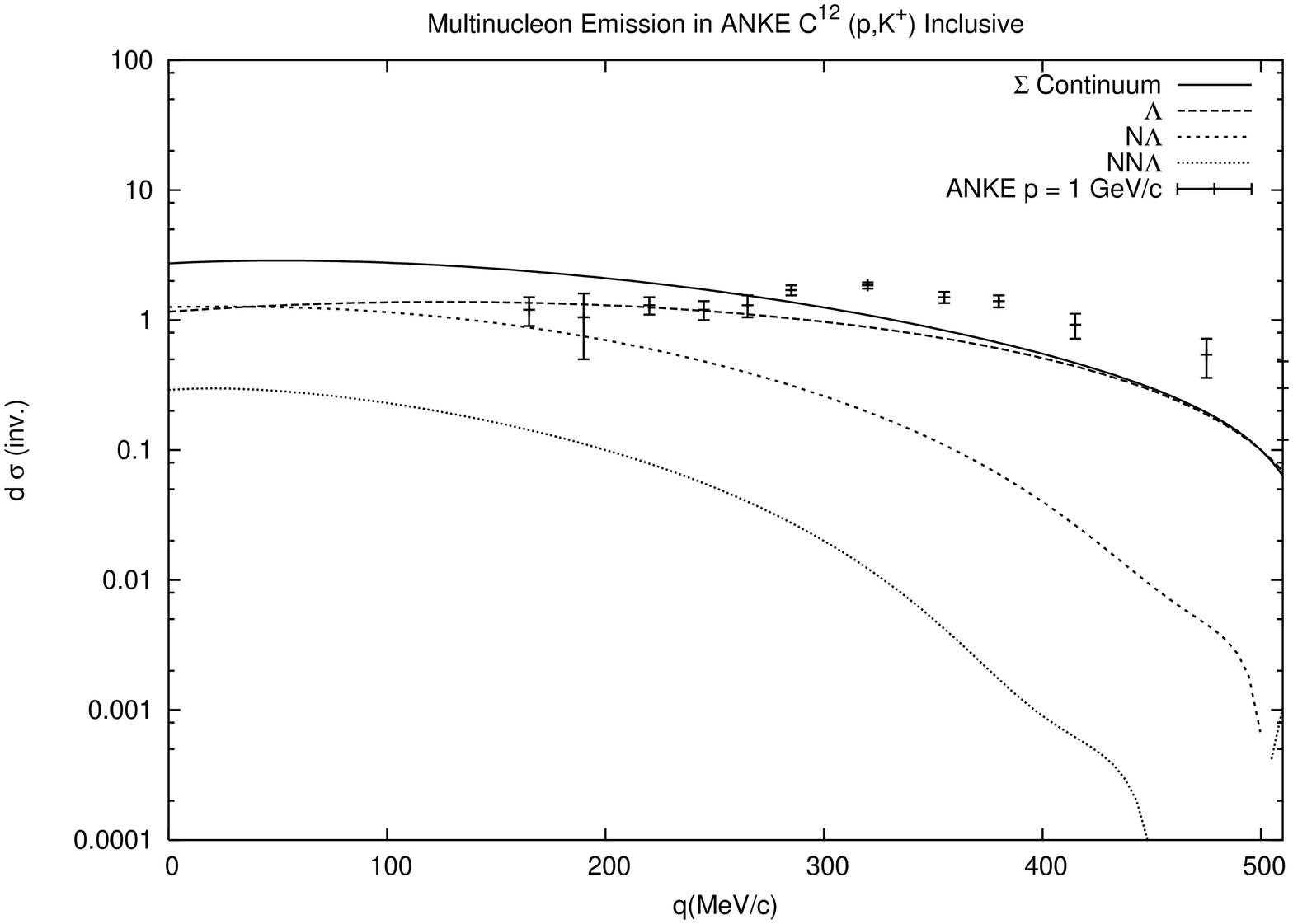}
\end{center}
\noindent {\footnotesize Fig. 1. Invariant cross section for $\mathrm{p+\, ^{12} C\rightarrow K^+
+X}$ momentum. The data are from ref.\cite{Koptev}}.
\end{figure}
We find that for low $K^+$ momenta $p_{K^+}\lesssim 200MeV/c$ the cross section is clearly
dominated by $\Lambda N$ and $\Lambda NN$ states in the continuum; more complicated final states
with 3 or more nucleons are already strongly suppressed. For increasing $K^+$ momenta $\gtrsim
400MeV/c$ the nucleon phase space cuts down multi nucleon emission and is clearly dominated by just
an unbound $\Lambda$ in the final state (together with the bound $^{12}C$ nucleus). It is clear,
though not shown in the calculation, that approaching exclusive $K^+$ production in $ ^{12}
\mathrm{C}(p,K^+)$, the formation of an unbound $K^+\Lambda$ continuum state is increasingly
suppressed up to the exclusive endpoint with the formation of a bound hypernucleus $
^{13}_{\Lambda} \mathrm{C}$.

\section{Summary and outlook}
In this short note we investigate the momentum distribution of $K^+$ in the inclusive production in proton induced scattering on $ ^{12} \mathrm{C}$. We find in the experimental momentum range of the $K^+$ with 100MeV/c $\lesssim p_{K^+} \lesssim $ 500MeV/c the transition of two to one nucleon in the continuum with increasing $K^+$ momenta. As expected, the structure of the cross section results from an interplay of the available phase space of the nucleons (and of the $\Lambda$) in the continuum, the strength of the $\pi, \rho$ induced rescattering mechanism and the momentum sharing among the bound nucleons.\\
\newline
\indent  At present our picture is preliminary and has to be substantiated by further data (most interesting would be the transition to (nearly) exclusive $K^+$ fusion\cite{Fetisov}) and by further calculations: so far our rescattering mechanism is too crude for quantitative predictions and for subtle information like on the $\pi$ or $K^+$ self energy\cite{Rudy,Buescher}. Thus attempts towards a covariant model calculation including full retardation, which allows for a free energy and momentum sharing among the interacting nucleons, would be highly desirable.


\begin{thebibliography}{0}
\bibitem{Machner} H. Machner, {\it nucl-ex/0505006};\\
T. Johansson et al., {\it Int. J. Mod. Phys.} {\bf A20} (2005) 1588.

\bibitem{Koptev} V. Koptev et al., {\it Phys. Rev. Lett.} {\bf 87} (2001) 022301.

\bibitem{Vries} H. de Vries, C. W. de Jager and C. de Vries, {\it At Data Nucl. Data Tables} {\bf 36} (1987) 495.

\bibitem{Groom} Part. Data Group (S. Eidelman et al.), {\it Phys. Lett.} {\bf B592} (2004) 1.

\bibitem{Sibirtsev} A. Sibirtsev, K. Tsushima and A. W. Thomas, {\it Phys. Lett.} {\bf B421} (1998) 59;\newline
K. Tsushima, A. Sibirtsev and A. W. Thomas, {\it Phys. Lett.} {\bf 390} (1997) 29.

\bibitem{Machleidt} R. Machleidt, {\it Adv. Nucl. Phys.} {\bf 19} (1989) 189.

\bibitem{Riska} D. O. Riska and G. E. Brown, {\it Nucl. Phys.} {\bf A679} (2001) 577.

\bibitem{Gradshteyn} I. S. Gradshteyn and I. M. Ryzhik, {\it Table of integrals, series and products S. 631} (1980) 716 {\it Academic Press, San Diego}.

\bibitem{Abramowitz} M. Abramowitz and I.A. Stegun, {\it Handbook of Math. Functions}\\ (Dover Publ. Inc., New York, 1970).

\bibitem{Shyam} R. Shyam and J. Knoll, {\it Nucl. Phys.} {\bf A426} (1984) 606.

\bibitem{Fetisov} V. N. Fetisov, {\it Nucl. Phys.} {\bf A639} (1998) 177;\\
R. Shyam, H. Lenske and U. Mosel, {\it Nucl. Phys.} {\bf A764} (2006) 313;\\
M. Dillig and M. Schott, {\it in preparation}.

\bibitem{Rudy} G. Z. Rudy et al., {\it Eur. Phys. J.} {\bf A15} (2002) 303.
\bibitem{Buescher} M. B\"uscher, {\it FZ J\"ulich} (2004) 1.

\end{thebibliography}
\end{document}